\documentclass[11pt]{article}

\usepackage[margin=1in]{geometry}
\usepackage{amsmath, amssymb}
\usepackage{graphicx}
\usepackage{booktabs}
\usepackage{tabularx}
\usepackage{array}
\usepackage{caption}
\usepackage{hanging}
\usepackage{setspace}
\usepackage{xcolor}
\usepackage[hidelinks]{hyperref}

\newcolumntype{Y}{>{\raggedright\arraybackslash}X}
\newcolumntype{Z}{>{\centering\arraybackslash}X}

\title{The Organizational Behavior of Agentic AI: Context, Boundaries, and Collective Intelligence in Human-Agent Workflows}

\author{Canhui Liu\\Department of Computer Science, University College London\\The AI Hub in Generative Models\\\texttt{canhui.liu@ucl.ac.uk}}

\date{\today}

\begin{document}

\maketitle

\begin{abstract}

Agentic artificial intelligence is increasingly deployed not as a single assistant but as a collective of planners, solvers, reviewers, memory managers, tool users, and orchestrators. These systems are entering organizational workflows under familiar labels such as teams, managers, committees, markets, and workflows. This article asks whether such agent collectives exhibit organizational behavior in a sense that is analytically comparable to, yet distinct from, human organizational behavior. I argue that agentic AI is a partial organizational analogue. It resembles human organization because it differentiates work, coordinates interdependence, performs recurrent routines, crosses boundaries, and produces collective outcomes. It differs because these patterns are not sustained by motivation, identity, trust, employment, socialization, or moral accountability. They are sustained by context architecture: prompts, memory, traces, schemas, tools, validators, and permissions. The article develops contextual transaction cost as the central mechanism linking these similarities and differences. Computational theorizing, synthetic task simulations, real LLM agent traces, and robustness analyses show that human-imitation forms often underperform when they add lossy handoffs, correlated deliberation, and verification burdens, whereas shared-state and adaptive forms perform better when they make context durable, inspectable, and task-contingent. The article contributes to organization studies by theorizing agentic AI as an emerging object of organizing and by specifying the interface conditions under which human and agentic organizational behavior can jointly support collective intelligence.

\end{abstract}


\section{Introduction}

Organization studies has often advanced when a new form of collective action disturbed an inherited category. The factory made division of labor and interdependence visible as organizational problems. The corporation made authority, employment, and administrative decision making central to the study of organizations. Networks, platforms, projects, and digital infrastructures later unsettled the assumption that organization is coextensive with the legally bounded firm. Agentic artificial intelligence now poses a related disturbance. AI is no longer only a decision aid, a predictive system, or a tool used by an employee. Increasingly, it appears as a set of computational agents arranged to perform work together.

This shift matters because agentic AI is already described in organizational terms. Designers speak of agent teams, manager agents, reviewer agents, legal review panels, research collectives, markets for task allocation, and societies of agents. The language is understandable. Agent collectives face familiar organizational problems. Work must be decomposed. Roles must be differentiated. Partial outputs must cross boundaries. Errors must be detected before they cascade. Evidence must be preserved across time. Tool permissions must be allocated. Accountability must be reconstructed from traces. These are not merely engineering concerns. They are problems of organizing.

Yet the same organizational language can mislead. Human organizations are not only information-processing systems. They are social orders shaped by employment relations, identity, trust, obligation, conflict, status, tacit knowledge, professional jurisdiction, and moral responsibility. AI agents do not enter organizations in this sense. They do not become socialized into roles, defend status, form informal coalitions, carry professional identity, or experience responsibility as a moral burden. They can be copied, reset, instrumented, and redeployed at little social cost. At the same time, they fail in ways that human analogies can hide. They lose context across handoffs, treat compressed summaries as authoritative, reproduce correlated errors, manufacture consensus, and depend heavily on prompt and protocol design.

The central question is therefore not whether agentic AI is an organization. That question forces a premature ontological answer. The more useful question is whether agentic AI exhibits organizational behavior, and if so, how that behavior is similar to and different from the organizational behavior of human collectives. This question is important because agentic AI is becoming embedded in human workflows. If agent collectives behaved like human teams, familiar organization forms could be transferred directly. If their behavior differs in systematic ways, then imitation may create costly dysfunction. Human-agent collaboration would require a theory of the boundary between human organizational behavior and agentic organizational behavior.

This article develops such a theory. I define the organizational behavior of agentic AI as the patterned, recurrent, and consequential behavior of AI agent collectives as they differentiate work, coordinate interdependence, process and preserve task context, and interact with human organizational workflows. The definition is behavioral rather than ontological. It does not claim that agent collectives are organizations in the same legal, sociological, or moral sense as firms, professions, communities, or agencies. It claims that agent collectives can display organized patterns of action that matter for organizational outcomes.

The article's argument has three parts. First, agentic AI collectives are functionally similar to human organizations because they face problems of differentiation, coordination, routines, boundaries, and collective knowledge. This similarity makes organization theory relevant. Second, agentic AI collectives are behaviorally different because their patterns of action are not sustained by human social embeddedness. They are sustained by context architecture: prompts, schemas, memory, traces, validators, tools, and permissions. This difference limits direct analogy to human teams, hierarchies, committees, and markets. Third, efficient human-agent collaboration depends on interface organizations that align human accountability with agentic context coordination.

The mechanism connecting these parts is contextual transaction cost. Coase (1937) and Williamson (1985) taught organization scholars to ask when one mode of coordination becomes more costly than another. Agentic AI requires a related but distinct cost concept. In agent collectives, the costly transaction is often the movement of context across boundaries. A planner's decomposition must be usable by a solver. A solver's claim must carry enough evidence for a reviewer. A reviewer must distinguish uncertainty from style. A shared memory system must preserve useful state without stabilizing error. A human manager must be able to audit the trace before the output enters an accountable workflow. These costs are not incidental. They determine when collective intelligence becomes productive and when it becomes wasteful.

The article supports the theory through computational theorizing. The empirical material includes a synthetic knowledge-work simulation of 8,000 tasks across seven agent organization forms, task-fixed-effects models comparing organization forms within the same task, scaling experiments that vary error correlation and communication mode, trace-instrumented real LLM runs across software repair, long-document question answering, legal review, and literature synthesis tasks, and robustness checks using open-source models, prompt variation, and a prompt-free reinforcement-learning organization selector. These materials are not presented as a population estimate of all agentic AI systems. They are used as mechanism-oriented evidence, consistent with simulation-based theory development in management research (Davis, Eisenhardt, \& Bingham, 2007; Harrison et al., 2007).

The contribution is theoretical. Agentic AI matters to organization studies not only because it automates or augments human work. It matters because it introduces organized nonhuman collectives into organizational life. Their behavior resembles human organizational behavior enough for organization theory to be useful, yet differs enough that unreflective analogy becomes dangerous. The article contributes to research on technology and work, routines and collective knowledge, algorithmic organizing, and collective intelligence by specifying the organizational behavior of agentic AI as a distinct object of analysis.

\section{Literature Review}

\subsection{Technology and the Reconfiguration of Organizing}

Organization studies has long moved beyond treating technology as an external instrument that enters an otherwise stable organization. Barley (1986) showed that technology can occasion new social orders at work. Orlikowski (2000) reframed technology use as the enactment of structures in practice. Later sociomaterial work sharpened this point by arguing that organizing is constituted through relations among people, artifacts, practices, and material arrangements rather than through social action alone (Orlikowski, 2007; Orlikowski \& Scott, 2008).

This literature matters for agentic AI because agents are not passive artifacts in a workflow. They generate plans, drafts, classifications, code patches, legal risk labels, summaries, and critiques. A planner agent's output can become a task structure. A reviewer agent's critique can become a decision criterion. A memory entry can become the durable version of what the system takes to be true. The prompt, context window, retrieval store, tool call, and trace are therefore not background infrastructure. They are part of the arrangement through which work is organized.

The best work on technology and organizing also warns against simple adoption narratives. Digital infrastructures can organize and disorganize at the same time (Ratner \& Plotnikof, 2022). Connection technologies can produce responsiveness while also intensifying obligation (Symon \& Pritchard, 2015). Platforms can appear novel while carrying older organizational logics forward (Steinberg, 2022). Agentic AI should be approached in the same way. The relevant question is not whether agents increase or reduce efficiency in general. It is how agentic arrangements reconfigure the conditions under which collective work is coordinated, remembered, evaluated, and made accountable.

This literature also clarifies what must be extended. Much sociomaterial research studies technologies that mediate human practice. Agentic AI adds computationally differentiated contributors that coordinate with one another before their outputs enter human practice. The internal organization of agents may be invisible unless the system is instrumented. Organization studies therefore needs concepts that can examine both the internal patterns of agent collectives and their embedding in human work.

\subsection{Collective Knowledge, Routines, and Boundary Work}

The second conversation concerns how organizations know and act collectively. Organizational knowledge is not simply the sum of individual knowledge. It is distributed across people, routines, artifacts, communities, and practices (Brown \& Duguid, 1991; Tsoukas, 1996). Hecker (2012) shows that collective knowledge can mean shared knowledge, complementary distributed knowledge, or knowledge embedded in artifacts and routines. This distinction is crucial for agentic AI. An agent collective may appear to know a task because the planner holds the decomposition, the retriever holds evidence identifiers, the solver holds a draft, the reviewer holds critique, and memory holds state. The collective result can exceed any component, but its knowledge is stored in prompts, traces, vector stores, logs, and schemas rather than in social practice.

Routines research provides a second anchor. Routines are recurrent patterns of interdependent action that stabilize work while allowing variation (Cohen \& Bacdayan, 1994; Feldman \& Pentland, 2003). Pentland, Haerem, and Hillison (2010) show that routines can be compared as patterned action. Agent traces make such patterns unusually visible. A trace may repeatedly show planning, retrieval, solving, verification, and synthesis. These patterns resemble routines because they recur and coordinate interdependence. They differ because they lack the human agency, memory, and situated repair through which human routines are enacted.

Boundary work is equally central. Knowledge boundaries require transfer, translation, and transformation as knowledge moves across groups, practices, and interests (Carlile, 2002, 2004). Boundary objects can travel across social worlds because they are both robust and plastic (Star \& Griesemer, 1989). Agent collectives face boundary problems whenever task context moves between agents, memory systems, tools, and human reviewers. The boundary is not only social. It is architectural. What is serialized? What is compressed? What evidence remains attached? What uncertainty is lost? What becomes durable? These questions determine whether a handoff preserves knowledge or creates drift.

Together, collective knowledge, routines, and boundary theory suggest that agentic AI should not be evaluated only by final answer quality. Its organizational behavior lies in the patterned movement of task context. A high-quality answer produced through opaque and fragile context movement may be difficult to embed in accountable work. Conversely, a slightly lower-quality answer with strong evidence preservation may be more useful in a legal, engineering, or scientific workflow.

\subsection{Algorithms, AI, and Human-Agent Work}

The third conversation concerns algorithms and AI in organizations. Research on algorithms at work has shown that digital systems increasingly allocate, evaluate, discipline, and coordinate work (Kellogg, Valentine, \& Christin, 2020). Learning algorithms alter not only tasks but also the organization of expertise and accountability (Faraj, Pachidi, \& Sayegh, 2018). AI creates an automation-augmentation paradox because it can both substitute for human action and complement human judgment (Raisch \& Krakowski, 2021). Studies of AI decision structures similarly show that delegation to AI changes how decisions are decomposed, reviewed, and governed (Shrestha, Ben-Menahem, \& von Krogh, 2019).

This literature has largely studied how algorithms organize humans or how humans collaborate with AI systems. Agentic AI adds a different unit of analysis. Before an AI output reaches a human workflow, multiple agents may already have planned, retrieved, debated, revised, verified, and synthesized. The internal organization of the AI system therefore shapes the human-AI relationship. A lawyer, engineer, or analyst does not merely interact with a model. They inherit the outcome of an agentic organization whose internal context boundaries may be strong or weak, visible or opaque, adaptive or fixed.

This shift also complicates collective intelligence research. Human collective intelligence depends on interaction patterns, diversity, participation, and coordination rather than individual ability alone (Malone, Laubacher, \& Dellarocas, 2010; Woolley et al., 2010). Agent collectives raise the same question in a different medium. Multiple agents can improve performance when they contribute independent search, complementary evidence, or effective verification. They can reduce performance when they multiply correlated errors or impose excessive context costs. The number of agents is therefore not the unit of explanation. The organization of context is.

\section{Theoretical Framework}

\subsection{Agentic AI as a Partial Organizational Analogue}

The theoretical premise of this article is that agentic AI is a partial organizational analogue. It is organizationally consequential, but not socially equivalent to human organization. The analogy is partial because agent collectives and human organizations share functional problems without sharing the same social foundations.

Five similarities make organization theory relevant. Agent collectives differentiate work into roles and subtasks. They coordinate interdependence among partial outputs. They enact recurrent routines visible in traces. They cross boundaries between agents, memory, tools, and human review points. They produce collective outcomes that cannot be reduced to a single agent's isolated response. These similarities connect agentic AI to classic concerns of bounded rationality, interdependence, information processing, routines, and coordination (March \& Simon, 1958; Thompson, 1967; Galbraith, 1973).

The differences are equally important. Human organizational behavior is grounded in motivation, identity, authority, trust, socialization, tacit knowledge, politics, and accountability. Agentic organizational behavior is grounded in context architecture. A reviewer agent does not enact professionalism because it belongs to a profession. It enacts review because a role, prompt, schema, tool permission, and validation rule call forth that behavior. A manager agent does not economize attention through social authority. It decomposes and synthesizes through context operations. A committee of agents does not create legitimacy through deliberation. It creates text that may or may not reflect independent evidence.

Figure 1 presents the theoretical architecture. Human workflows impose demands for judgment, accountability, continuity, and legitimacy. Agentic AI collectives generate differentiated and recurrent computational action. Their productive connection depends on interface organizations that translate between human accountability and agentic context coordination.

\begin{figure}[htbp]
\centering
\includegraphics[width=0.92\textwidth]{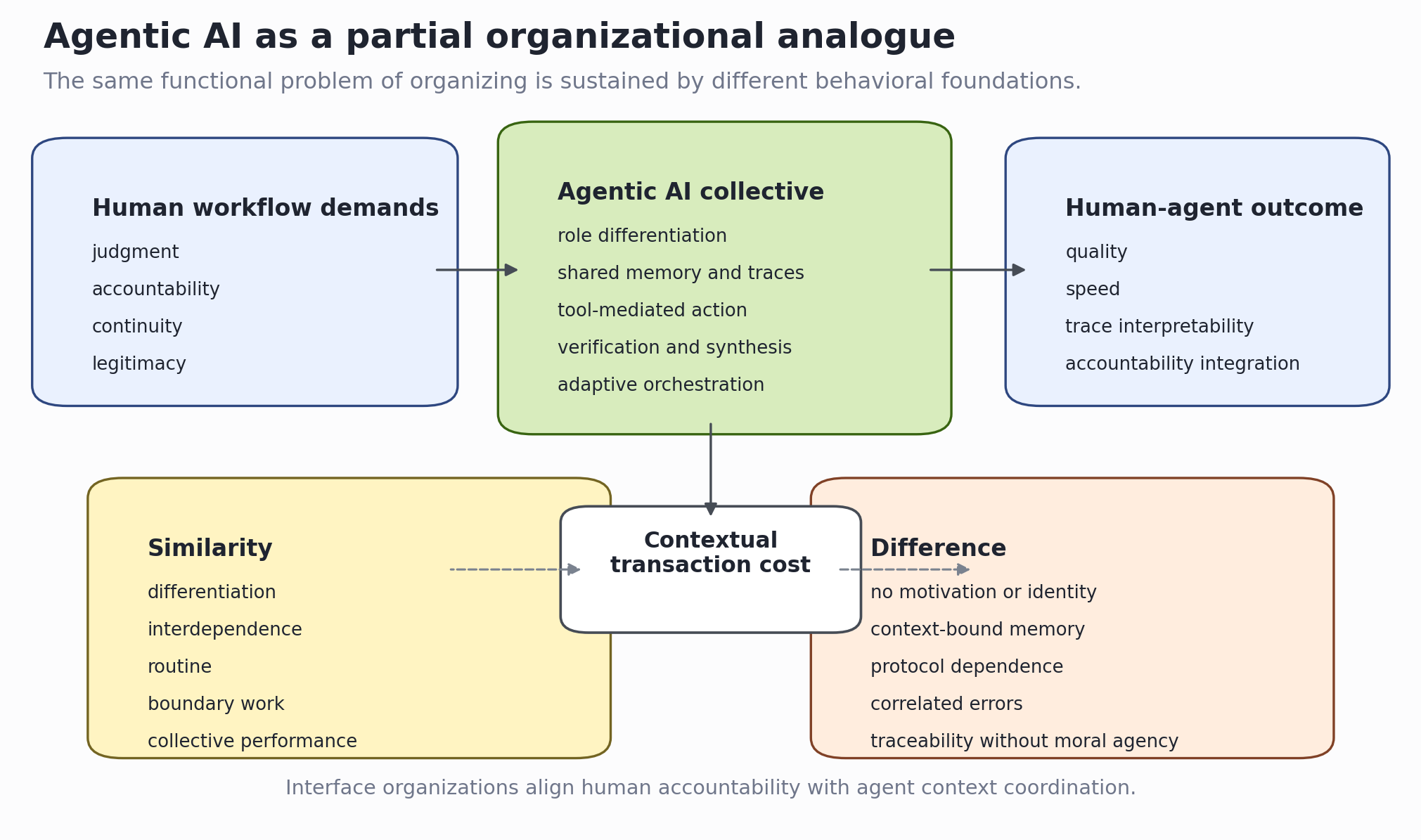}
\caption{Agentic AI as a partial organizational analogue}
\end{figure}

The partial-analogue framing avoids two errors. The first is instrumentalism, which treats agentic AI as a neutral tool. The second is anthropomorphism, which treats agent collectives as if they were human teams. Agentic AI is neither. It is an object of organizing whose behavior must be explained in its own medium.

\subsection{Contextual Transaction Cost}

Contextual transaction cost is the cost of making task context usable across boundaries in an agent collective. The concept extends transaction cost reasoning, but the transaction is different. In classic economic theories of organization, the relevant costs concern search, bargaining, contracting, monitoring, enforcement, and opportunism (Coase, 1937; Williamson, 1985). In agentic AI, the immediate cost often concerns context transfer, compression, interpretation, verification, and governance.

At the event level, contextual transaction cost can be represented as

\[
CTC_e =
w_\tau \tau_e +
w_h H_e +
w_c C_e +
w_d D_e +
w_v V_e +
w_g G_e ,
\]

where \(\tau_e\) is token and latency burden, \(H_e\) is a cross-agent handoff, \(C_e\) is compression loss, \(D_e\) is semantic drift, \(V_e\) is verification burden, and \(G_e\) is governance cost. At the run level,

\[
CTC_{to}=\sum_{e \in E(t,o)} CTC_e ,
\]

where \(t\) indexes the task, \(o\) indexes the organization form, and \(E(t,o)\) is the trace generated by organization \(o\) on task \(t\).

The value of an agentic organization can then be decomposed as

\[
V(O,T)=
G_s(O,T)+G_p(O,T)+G_d(O,T)+G_v(O,T)
-CTC(O,T)-K(O,T)-R_g(O,T).
\]

The gain terms refer to specialization, parallelism, diversity, and verification. \(K\) denotes ordinary compute cost and \(R_g\) denotes governance risk. The expression is not intended as a universal production function. It specifies the theoretical forces that determine whether collective agent behavior becomes intelligent, redundant, or hazardous.

The empirical outcome used in the studies is collective efficiency:

\[
CE_{to}=
\frac{Q_{to}\times P(S_{to}=1)}
{1+\lambda Cost_{to}},
\]

where \(Q_{to}\) is task quality, \(P(S_{to}=1)\) is success probability, and \(Cost_{to}\) includes compute, verification, management, and contextual transaction costs. This measure is deliberately stricter than quality. In human workflows, an output that is marginally better but difficult to audit, slow to produce, or expensive to verify may be a worse organizational outcome.

The framework yields a simple inequality:

\[
\Delta G_s+\Delta G_p+\Delta G_d+\Delta G_v >
\Delta CTC+\Delta K+\Delta R_g .
\]

Agentic collective organization improves performance when the marginal gains from specialization, parallel search, diversity, and verification exceed the marginal costs of context transactions, compute, and governance risk. It damages performance when the boundaries it creates are more costly than the gains they enable.

\subsection{Interface Organizations}

The framework culminates in the concept of an interface organization. An interface organization is the designed arrangement that connects human organizational behavior with agentic organizational behavior. It specifies which agent outputs can enter human workflows, what evidence must travel with them, what uncertainty must be disclosed, what permissions agents have, what traces are preserved, and where human judgment remains mandatory.

This concept matters because agentic AI will usually be embedded in existing human organizations rather than replacing them wholesale. A law firm may use agents to review contracts, but lawyers remain accountable for advice. A software team may use coding agents, but engineers remain accountable for deployment. A research group may use literature agents, but authors remain accountable for claims. In each case, the design problem is not only model quality. It is the alignment between agentic context coordination and human accountability.

Table 1 summarizes the mechanism expectations that guide the empirical design.

\begin{table}[htbp]
\centering
\small
\caption{Mechanism expectations for the organizational behavior of agentic AI}
\begin{tabularx}{\textwidth}{@{}YYY@{}}
\toprule
Mechanism & Theoretical claim & Empirical implication \\
\midrule
Functional similarity & Agent collectives face differentiation, coordination, routine, boundary, and collective knowledge problems. & Multi-agent forms can outperform compact execution when tasks are decomposable, verifiable, and diverse. \\
Behavioral difference & Agentic behavior is sustained by context architecture rather than human motivation, identity, and trust. & Human-imitation forms underperform when they add handoffs without preserving evidence or reducing drift. \\
Contextual transaction cost & Collective gains require context to remain usable across boundaries. & Efficiency declines when handoffs, compression loss, semantic drift, and verification burden accumulate. \\
Pseudo-diversity & More agents do not imply independent judgment when errors are correlated. & Free-text debate scales poorly unless agents draw on independent models, tools, or evidence. \\
Interface organization & Human-agent collaboration requires alignment between traces, evidence, permissions, and accountability. & Agent-native forms become more valuable when they expose evidence and adapt coordination to task conditions. \\
\bottomrule
\end{tabularx}
\end{table}

\section{Research Design and Methodology}

\subsection{Computational Theorizing Strategy}

The empirical strategy is computational theorizing. The aim is to formalize a mechanism, vary the conditions under which it operates, and observe whether the resulting patterns sharpen the theoretical claim. Simulation and trace analysis are appropriate here because agent collectives can be instrumented at a level of detail rarely available in human organizations. Every prompt, handoff, memory write, tool call, validation event, and output can be logged. This makes it possible to study organizational behavior as patterned action rather than as a metaphor.

The design combines three sources of evidence. Study 1 uses a synthetic task landscape to compare seven agent organization forms across identical tasks. Study 2 uses trace-instrumented real LLM runs to test whether contextual transaction costs can be estimated in actual agent executions. Study 3, reported in the online appendix, examines robustness through prompt variation, open-source model runs, and a prompt-free fitted DQN organization selector. The studies are summarized in Table 2.

\begin{table}[htbp]
\centering
\small
\caption{Research design}
\begin{tabularx}{\textwidth}{@{}YZZZ@{}}
\toprule
Study & Evidence & Main purpose & Main manuscript role \\
\midrule
Study 1 & 8,000 synthetic knowledge-work tasks evaluated under seven organization forms & Identify task-organization patterns under controlled variation & Main evidence \\
Study 2 & Real LLM traces across software repair, long-document QA, legal review, and literature synthesis & Test whether CTC can be observed in actual agent traces & Main evidence \\
Study 3 & Prompt robustness, open-source model comparisons, and prompt-free DQN selector & Address robustness and prompt-dependence objections & Appendix evidence \\
\bottomrule
\end{tabularx}
\end{table}

This design does not substitute for field studies of human-agent workflows. It prepares them. It identifies the internal organizational patterns that field studies should examine: context boundaries, evidence flows, recurrent trace routines, verification points, and accountability interfaces.

\subsection{Study 1: Synthetic Knowledge-Work Simulation}

Study 1 generates 8,000 synthetic knowledge-work tasks. Each task is described by seven dimensions: complexity, decomposability, coupling, verifiability, ambiguity, risk, and knowledge velocity. These dimensions are organizational rather than merely technical. Decomposability creates potential specialization gains. Complexity creates potential parallelism gains. Coupling and ambiguity increase contextual transaction costs. Verifiability creates potential verification gains. Risk increases governance cost. Knowledge velocity increases the danger of stale or incomplete context.

Each task is evaluated under seven organization forms. The single expert represents compact execution with no internal handoff. The pipeline standard operating procedure represents sequential specialization. The hierarchy manager represents central decomposition and synthesis. The committee debate represents parallel free-text deliberation. The market contract form represents structured decentralized allocation. The blackboard memory form represents coordination through shared task and evidence state. The adaptive meta-organization represents a CTC-aware orchestrator that changes topology, communication medium, validation intensity, and permission scope based on task conditions.

The primary outcome is collective efficiency. The analysis also reports quality, success, cost, CTC share, semantic drift, and governance risk. Because each task is evaluated under every organization form, the main models estimate within-task differences:

\[
CE_{to}=\alpha_t+\sum_{o \neq single}\beta_o Org_{o,to}+\varepsilon_{to}.
\]

The task fixed effect \(\alpha_t\) absorbs all task-level characteristics. The coefficients \(\beta_o\) compare each organization form with the single expert on the same tasks. Standard errors are clustered by task. Parallel models use quality and success probability as outcomes.

\subsection{Study 2: Trace-Instrumented Real LLM Benchmark}

Study 2 uses real LLM agent traces. The benchmark contains four task families modeled after widely used AI evaluation settings: SWE-bench-style software repair, LongBench-style long-document question answering, LegalBench-style legal review, and QASPER/PaperQA-style literature synthesis (Dasigi et al., 2021; Guha et al., 2023; Bai et al., 2024; Jimenez et al., 2024). The current implementation uses local executable and automatically scored tasks rather than official benchmark splits. The purpose is to validate trace instrumentation and CTC estimation across real model calls, not to produce a public leaderboard.

Each task is executed under four organization forms: single, pipeline, blackboard, and adaptive. The single form asks one agent to solve the task. The pipeline form moves through planning, solving, review, and synthesis. The blackboard form uses shared evidence memory. The adaptive form selects compact, contract-based, shared-state, or high-verification structures based on task descriptors.

Each event trace records task, domain, organization, agent role, action type, input and output token estimates, evidence identifiers, verification result, latency, handoffs, and CTC components. This event-level design is central to the theory. It makes organization visible as patterned action. The analysis can ask not only whether the answer was correct, but how the collective moved context, preserved evidence, crossed boundaries, and incurred verification costs.

The real LLM runs include OpenAI \texttt{gpt-4.1-mini}, Gemini \texttt{gemini-2.5-flash}, Anthropic \texttt{claude-haiku-4-5-20251001}, and two local open-source models served through Ollama: \texttt{qwen2:7b-instruct} and \texttt{mistral:latest}. Quality is scored through task-specific evaluators. Success is a thresholded indicator of adequate task solution.

\section{Empirical Evidence}

\subsection{Human-Imitation Forms Do Not Necessarily Produce Human-Like Coordination}

The first empirical pattern is that visible organizational form is a poor guide to agentic organizational behavior. Figure 2 reports simulated collective efficiency by organization form and by CTC quartile. Adaptive meta-organization and blackboard memory are the strongest forms. The single expert remains competitive because it avoids internal context transactions. Pipeline, hierarchy, and committee forms perform poorly when their handoffs and deliberation create more context cost than collective gain.

\begin{figure}[htbp]
\centering
\includegraphics[width=0.92\textwidth]{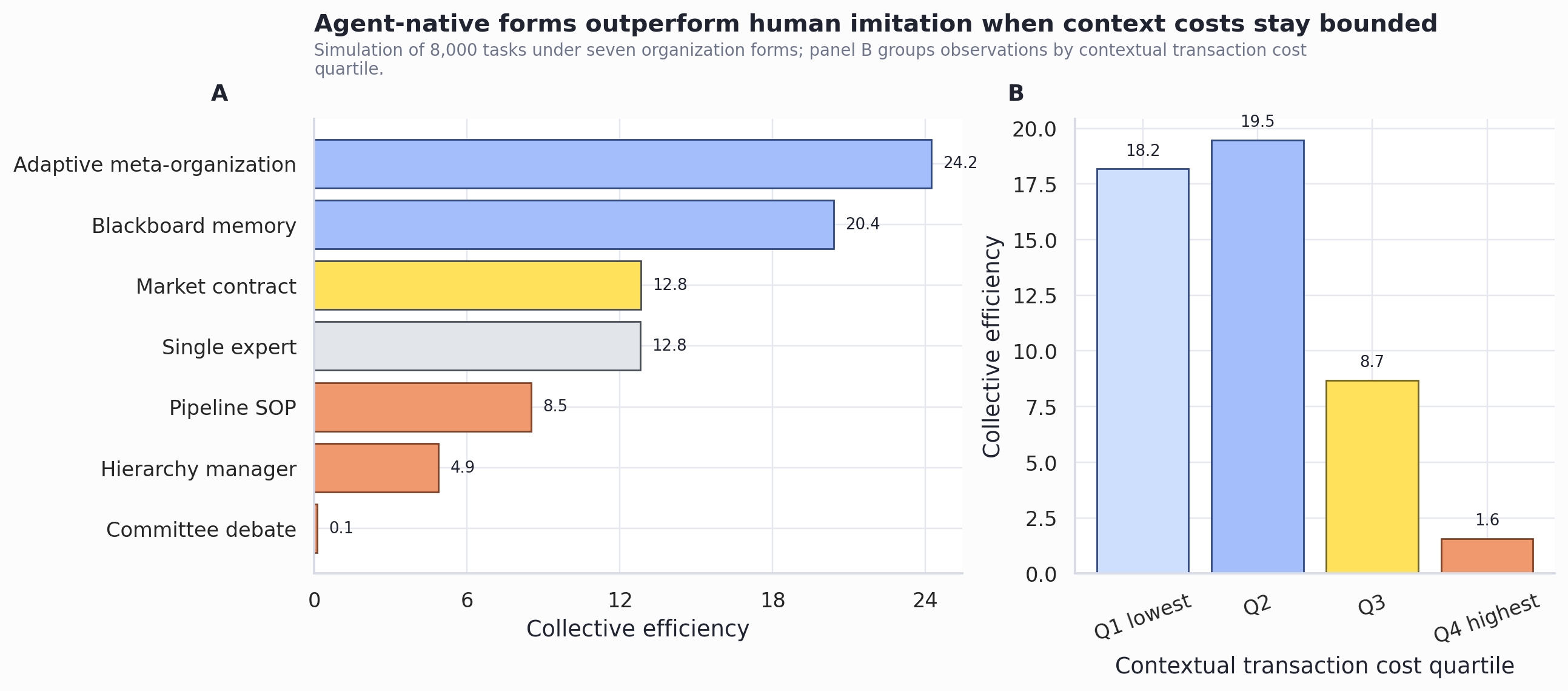}
\caption{Organization form and contextual transaction cost in the simulation}
\end{figure}

The family-level contrast is large. Agent-native forms are 395.26\% more efficient than human-imitation forms. Adaptive meta-organization improves collective efficiency by 89.24\% relative to the single expert. Blackboard memory is 139.44\% more efficient than the best human-imitation form. These differences are not simply artifacts of task allocation because each organization form is evaluated on the same tasks.

Table 3 reports task-fixed-effects estimates. Compared with the single expert, adaptive meta-organization increases collective efficiency by 11.43 points, quality by 14.00 points, and success probability by 23.24 percentage points. Blackboard memory increases collective efficiency by 7.60 points, quality by 11.74 points, and success probability by 19.48 percentage points. Pipeline, hierarchy, and committee forms are negative across the main outcomes. The committee result is especially instructive. It looks like a familiar human deliberative form, but in the simulation it creates high CTC and low independent epistemic gain.

\begin{table}[htbp]
\centering
\small
\caption{Task-fixed-effects estimates comparing organization forms with the single expert}
\begin{tabularx}{\textwidth}{@{}YZZZ@{}}
\toprule
Organization form & Collective efficiency & Quality & Success probability \\
\midrule
Pipeline SOP & -4.28*** (0.06) & -6.17*** (0.11) & -0.053*** (0.001) \\
Hierarchy manager & -7.92*** (0.07) & -11.49*** (0.10) & -0.116*** (0.001) \\
Committee debate & -12.69*** (0.09) & -42.10*** (0.09) & -0.240*** (0.002) \\
Market contract & 0.03 (0.07) & 3.33*** (0.10) & 0.058*** (0.002) \\
Blackboard memory & 7.60*** (0.09) & 11.74*** (0.10) & 0.195*** (0.002) \\
Adaptive meta-organization & 11.43*** (0.10) & 14.00*** (0.11) & 0.232*** (0.002) \\
\bottomrule
\end{tabularx}
\end{table}

Notes: N = 56,000 task-organization observations. All models include task fixed effects. Standard errors clustered by task are in parentheses. The omitted category is single expert. *** p < .001.

This finding matters theoretically because it separates similarity of form from similarity of behavior. A hierarchy can economize on attention and authority in human organizations because it rests on role expectations, employment relations, and accountability. In an agent collective, the same visible form may become repeated context compression. A committee can pool perspectives and confer legitimacy in human groups. In an agent collective, it may produce correlated paraphrases without independent evidence. A pipeline can stabilize routine human work. In an agent collective, it can propagate early errors if intermediate representations are not independently checked.

The implication is not that hierarchy, committees, or pipelines are always bad. It is that their value depends on whether they solve the context problem of agentic AI. Human-like forms become useful only when redesigned as context architectures rather than copied as social forms.

\subsection{Contextual Transaction Cost Explains When Collective Intelligence Becomes Efficient}

The second empirical pattern concerns the mechanism. Figure 2 also shows that collective efficiency is highest at moderate levels of contextual transaction cost and collapses at the highest quartile. The lowest CTC quartile has mean efficiency of 18.18. The second quartile has mean efficiency of 19.46. The third quartile falls to 8.68. The highest quartile falls to 1.56. Moving from the lowest to the highest CTC quartile corresponds to a 91.41\% decline in efficiency.

This pattern clarifies why the single expert remains competitive but not dominant. Compact execution avoids CTC but gives up specialization and verification gains. Moderate CTC can be productive because it reflects useful decomposition, evidence sharing, and review. Excessive CTC is destructive because it reflects repeated handoffs, lossy compression, semantic drift, and verification burden.

Blackboard memory and adaptive orchestration illustrate agent-native organization. Blackboard memory changes communication from serial message passing to shared state. It creates durable task, evidence, and critique objects that reduce repeated reconstruction. This resembles a boundary object, but the object is machine-readable, schema-constrained, and traceable. Its weakness is also architectural. If shared memory stabilizes an error, the error can become the collective's working reality.

Adaptive meta-organization treats organization form as a variable rather than a chart. It selects compact execution when coordination costs exceed gains, shared memory when evidence reuse matters, structured contracts when modular subtasks must be allocated, and stronger verification when risk and ambiguity are high. This result is consistent with contingency thinking, but the relevant contingency is context architecture fit. The central question is whether the form of context movement matches decomposability, coupling, verifiability, ambiguity, risk, and velocity.

\subsection{Scale Requires Engineered Diversity and Structured Communication}

The third empirical pattern concerns the size of agent collectives. Figure 3 reports the efficient number of agents under different communication modes and error-correlation conditions. Free-text debate has an optimal scale of one agent across high, medium, and low error correlation. Structured blackboard communication supports larger collectives when error correlation is low. Contract-latent hybrid communication supports the largest efficient collectives, reaching five to ten agents depending on error correlation.

\begin{figure}[htbp]
\centering
\includegraphics[width=0.92\textwidth]{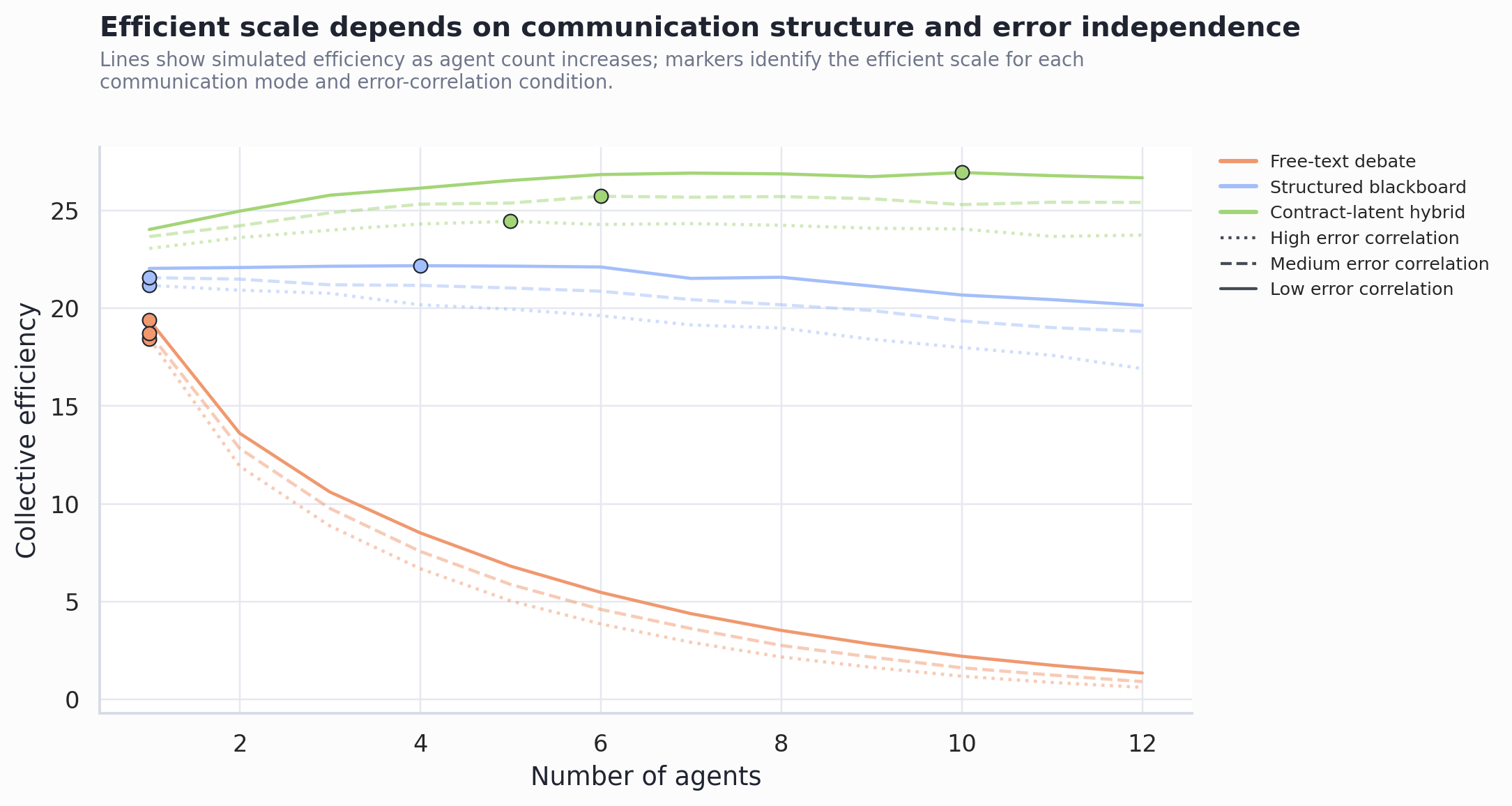}
\caption{Efficient agent scale depends on communication structure and error correlation}
\end{figure}

The interpretation is straightforward. Additional agents help when they contribute independent information, search paths, tools, or verification capacity. They hurt when they share the same error structure and communicate through lossy natural language. A five-agent system using the same model, prompt template, and retrieval context may be less diverse than a two-agent system using different models and independent evidence. Agent diversity is therefore not a headcount property. It is an engineered condition.

This finding extends collective intelligence theory. Human groups can benefit from diversity rooted in experience, identity, expertise, and social position. Agent collectives require different sources of diversity: model heterogeneity, tool heterogeneity, independent retrieval, adversarial verification, and different evidence sources. Without such independence, multi-agent organization can create pseudo-diversity. It gives the appearance of deliberation without the epistemic benefits of independent judgment.

\subsection{Real LLM Traces Reveal the Boundary Condition for Human-Agent Embedding}

The real LLM traces complicate any simple claim that more organization is always better. Figure 4 plots organization forms by mean quality and CTC-adjusted efficiency. Commercial models show high single-agent efficiency because compact execution avoids coordination overhead. Blackboard and adaptive forms sometimes improve quality, but the improvement comes with higher CTC. Open-source models show the same tradeoff with lower baseline competence and greater fragility.

\begin{figure}[htbp]
\centering
\includegraphics[width=0.92\textwidth]{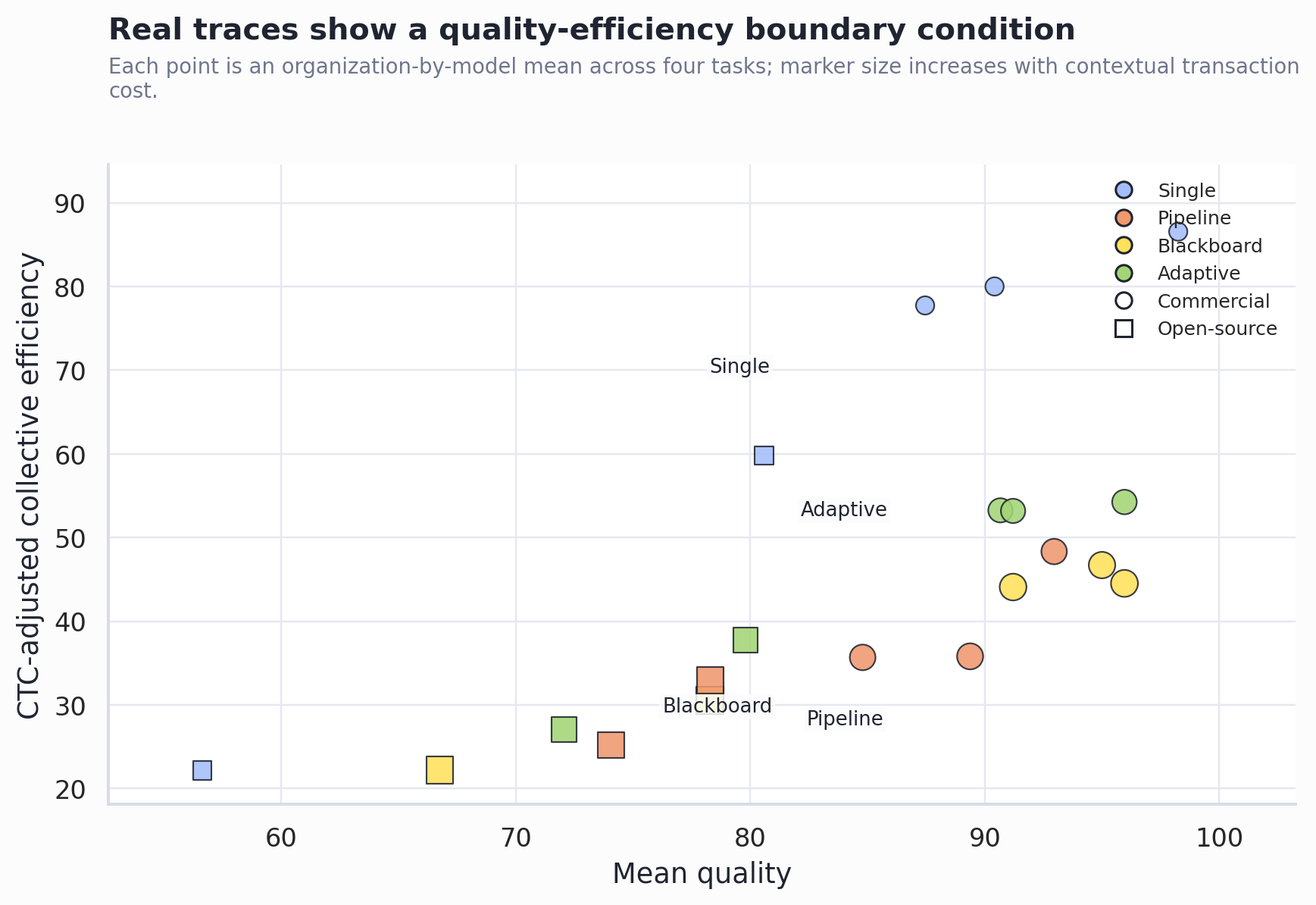}
\caption{Real LLM trace evidence on quality and CTC-adjusted efficiency}
\end{figure}

Table 4 summarizes the trace evidence. In the commercial runs, single execution has mean quality of 92.04, success of 100\%, CTC index of 0.93, and collective efficiency of 81.44. Blackboard has the highest mean quality at 94.06 but lower efficiency at 45.11 because its CTC index rises to 9.79. Adaptive organization has mean quality of 92.61 and efficiency of 53.57. In the open-source runs, single execution again has the highest efficiency because of low coordination cost, while adaptive and pipeline improve quality and success at higher CTC.

\begin{table}[htbp]
\centering
\small
\caption{Real LLM trace summary}
\begin{tabularx}{\textwidth}{@{}YZZZZZZZ@{}}
\toprule
Model set & Organization & Runs & Quality & Success & Handoffs & CTC index & Collective efficiency \\
\midrule
Commercial & Single & 12 & 92.04 & 1.00 & 0.00 & 0.93 & 81.44 \\
Commercial & Pipeline & 12 & 89.04 & 0.83 & 3.00 & 8.57 & 39.94 \\
Commercial & Blackboard & 12 & 94.06 & 1.00 & 4.00 & 9.79 & 45.11 \\
Commercial & Adaptive & 12 & 92.61 & 1.00 & 2.75 & 6.99 & 53.57 \\
Open-source & Single & 8 & 68.60 & 0.50 & 0.00 & 1.09 & 40.98 \\
Open-source & Pipeline & 8 & 76.18 & 0.63 & 3.00 & 8.92 & 29.06 \\
Open-source & Blackboard & 8 & 72.53 & 0.63 & 4.00 & 10.28 & 26.39 \\
Open-source & Adaptive & 8 & 75.92 & 0.63 & 2.75 & 7.27 & 32.42 \\
\bottomrule
\end{tabularx}
\end{table}

Notes: Commercial runs include OpenAI \texttt{gpt-4.1-mini}, Gemini \texttt{gemini-2.5-flash}, and Anthropic \texttt{claude-haiku-4-5-20251001}. Open-source runs include Qwen2 7B instruct and Mistral served locally through Ollama. Tasks cover software repair, long-document QA, legal review, and literature synthesis.

The trace evidence defines a boundary condition. When tasks are small, low risk, and within the competence of a strong model, a single agent may be the most efficient organization. When tasks are decomposable, high risk, evidence-intensive, or require independent verification, multi-agent forms can be justified despite higher CTC. The correct design question is therefore not whether agentic AI should be single-agent or multi-agent. It is which context boundaries are worth creating for the task and how those boundaries will remain auditable for human work.

\section{Discussion}

The evidence supports the article's central claim: agentic AI collectives exhibit organizational behavior, but they are not human organizations in miniature. They differentiate work, coordinate interdependence, enact routines, cross boundaries, and generate collective outcomes. These similarities make organization theory indispensable. Yet the mechanisms that sustain the behavior differ. Agentic organizational behavior is produced through prompts, context windows, shared state, traces, validators, tools, and permissions rather than through motivation, identity, trust, socialization, and moral accountability.

This distinction is the theoretical contribution. If agentic AI were simply a tool, organization studies could study adoption, use, and resistance. If it were simply a human-like organization, existing organization theory could be transferred directly. Agentic AI is neither. It is a partial organizational analogue: familiar at the level of functional problems, unfamiliar at the level of behavioral foundations.

Contextual transaction cost explains why this distinction matters. In human organizations, context often travels through tacit knowledge, shared history, informal correction, and responsibility norms. In agent collectives, context must be serialized, retrieved, compressed, validated, and made auditable. Coordination fails when these operations become more costly than the collective gains they enable. This is why visible organizational forms can mislead. A hierarchy, committee, pipeline, market, or shared-memory system must be evaluated by how it moves context, preserves evidence, and supports accountability.

The article contributes to four conversations. It extends sociomaterial approaches to technology at work by theorizing the internal organization of agentic AI as part of the sociomaterial arrangement of work. It contributes to routines and collective knowledge research by treating agent traces as recurrent action patterns and by showing how collective knowledge can be distributed through machine-readable state. It extends algorithmic management and AI-in-organization research by shifting attention from how algorithms organize humans to how AI agents are organized before they enter human workflows. It contributes to collective intelligence research by showing that multi-agent intelligence depends on context architecture, error independence, and verification, not on the number of agents alone.

The practical implication is that human-agent collaboration requires interface organizations. It is not enough to add a reviewer agent, because review may share the original error. It is not enough to add a manager agent, because management may become lossy summarization. It is not enough to add memory, because memory may stabilize mistakes. It is not enough to create a committee, because debate may create pseudo-diversity. The design question is comparative: what collective gain does this boundary create, what contextual transaction cost does it impose, and how does the trace support human accountability?

\section{Implications for Organization Theory and Design}

For organization theory, agentic AI shows that organizational behavior can be studied beyond human actors without erasing the distinctiveness of human sociality. The point is not to grant machines social personhood. It is to distinguish functional organization problems from social organization foundations. Differentiation, coordination, routines, boundaries, and collective performance can exist in agent collectives. Motivation, identity, trust, and moral accountability cannot be assumed.

For research on technology and work, the article suggests a new unit of analysis: the human-agent workflow. This unit includes human roles, agent roles, prompts, traces, tools, memory, evidence stores, and accountability systems. Field studies should examine how workers interpret agent traces, when they trust or override agent collectives, how responsibility is assigned after agent error, and how professional jurisdiction changes when agents perform intermediate work.

For organization design, the article suggests that managers should evaluate agentic AI through context boundaries rather than organizational metaphors. When a vendor describes an AI team, the relevant questions are concrete. What work is differentiated? What evidence travels between agents? Where is context lost? How are errors detected? What prompts and schemas stabilize behavior? Which human actor remains accountable? The answer may be a single agent, a shared-memory system, an adaptive orchestrator, or a human-led workflow with narrow agent support.

For computer science, the article suggests a technical path toward collective intelligence. Better agent collectives require lower contextual transaction costs, stronger trace validators, typed handoff schemas, uncertainty-preserving memory, genuine error diversity, and learned organization policies. Organization theory supplies the constructs: interdependence, routines, boundaries, accountability, and CTC. Reinforcement learning and bandit methods can help systems learn when each organization form is worth its cost.

\section{Conclusion}

Agentic AI is becoming collective and organizationally embedded. The key question is not whether these systems are organizations in the same sense as human organizations. They are not. The key question is whether they exhibit organizational behavior that matters for human work. This article argues that they do. Agent collectives differentiate work, coordinate interdependence, enact routines, cross boundaries, and produce collective outcomes. Yet their behavior is grounded in context, protocol, memory, tools, traces, and validation rather than human motivation, identity, trust, and moral agency.

This similarity and difference has practical consequences. Human organizations should not simply import familiar organizational charts into agent systems. They should design interface organizations that align human accountability with agentic context coordination. The efficiency of collective intelligence depends on whether the gains from specialization, diversity, parallelism, and verification exceed contextual transaction costs. The future of human-agent collaboration will therefore be shaped not only by better models, but by better organization theory for agentic AI.

\section*{References}

\begin{hangparas}{0.25in}{1}

Alchian, A. A., \& Demsetz, H. (1972). Production, information costs, and economic organization. American Economic Review, 62(5), 777-795.\\

Arrow, K. J. (1974). The Limits of Organization. Norton.\\

Bai, Y., Lv, X., Zhang, J., Lyu, H., Tang, J., Huang, Z., Du, Z., Liu, X., Zeng, A., Hou, L., Dong, Y., Tang, J., \& Li, J. (2024). LongBench: A bilingual, multitask benchmark for long context understanding. Proceedings of the Association for Computational Linguistics.\\

Barley, S. R. (1986). Technology as an occasion for structuring: Evidence from observations of CT scanners and the social order of radiology departments. Administrative Science Quarterly, 31(1), 78-108.\\

Benkler, Y. (2002). Coase's Penguin, or, Linux and the nature of the firm. Yale Law Journal, 112(3), 369-446.\\

Brown, J. S., \& Duguid, P. (1991). Organizational learning and communities-of-practice: Toward a unified view of working, learning, and innovation. Organization Science, 2(1), 40-57.\\

Brynjolfsson, E., Li, D., \& Raymond, L. R. (2025). Generative AI at work. Quarterly Journal of Economics, 140(2), 889-942.\\

Carlile, P. R. (2002). A pragmatic view of knowledge and boundaries: Boundary objects in new product development. Organization Science, 13(4), 442-455.\\

Carlile, P. R. (2004). Transferring, translating, and transforming: An integrative framework for managing knowledge across boundaries. Organization Science, 15(5), 555-568.\\

Coase, R. H. (1937). The nature of the firm. Economica, 4(16), 386-405.\\

Cohen, M. D., \& Bacdayan, P. (1994). Organizational routines are stored as procedural memory: Evidence from a laboratory study. Organization Science, 5(4), 554-568.\\

Cyert, R. M., \& March, J. G. (1963). A Behavioral Theory of the Firm. Prentice-Hall.\\

Dasigi, P., Lo, K., Beltagy, I., Cohan, A., Smith, N. A., \& Gardner, M. (2021). A dataset of information-seeking questions and answers anchored in research papers. Proceedings of NAACL.\\

Davis, J. P., Eisenhardt, K. M., \& Bingham, C. B. (2007). Developing theory through simulation methods. Academy of Management Review, 32(2), 480-499.\\

Faraj, S., Pachidi, S., \& Sayegh, K. (2018). Working and organizing in the age of the learning algorithm. Information and Organization, 28(1), 62-70.\\

Feldman, M. S., \& Pentland, B. T. (2003). Reconceptualizing organizational routines as a source of flexibility and change. Administrative Science Quarterly, 48(1), 94-118.\\

Galbraith, J. R. (1973). Designing Complex Organizations. Addison-Wesley.\\

Guha, N., Nyarko, J., Ho, D. E., Re, C., Chilton, A., Narayana, A., Chohlas-Wood, A., Peters, A., Waldon, B., Rockmore, D. N., Zambrano, D., Talisman, J., Hoque, E., Surani, F., Fagan, F., Sarfaty, G., Dickinson, G. M., Porat, H., Hegland, J., Wu, J., Nudell, J., Niklaus, J., Nay, J. J., Choi, J. H., Tobia, K., Hagan, M., Ma, M., Livermore, M. A., Rasumov-Rahe, N., Holzenberger, N., Kolt, N., Henderson, P., Rehaag, S., Goel, S., Gao, S., Williams, S., Gandhi, S., Zur, T., Iyer, V., \& Li, Z. (2023). LegalBench: A collaboratively built benchmark for measuring legal reasoning in large language models. Advances in Neural Information Processing Systems.\\

Harrison, J. R., Lin, Z., Carroll, G. R., \& Carley, K. M. (2007). Simulation modeling in organizational and management research. Academy of Management Review, 32(4), 1229-1245.\\

Hecker, A. (2012). Knowledge beyond the individual? Making sense of a notion of collective knowledge in organization theory. Organization Studies, 33(3), 423-445.\\

Hong, L., \& Page, S. E. (2004). Groups of diverse problem solvers can outperform groups of high-ability problem solvers. Proceedings of the National Academy of Sciences, 101(46), 16385-16389.\\

Jimenez, C. E., Yang, J., Wettig, A., Yao, S., Pei, K., Press, O., \& Narasimhan, K. (2024). SWE-bench: Can language models resolve real-world GitHub issues? International Conference on Learning Representations.\\

Kellogg, K. C., Valentine, M. A., \& Christin, A. (2020). Algorithms at work: The new contested terrain of control. Academy of Management Annals, 14(1), 366-410.\\

Malone, T. W., Laubacher, R., \& Dellarocas, C. (2010). The collective intelligence genome. MIT Sloan Management Review, 51(3), 21-31.\\

March, J. G., \& Simon, H. A. (1958). Organizations. Wiley.\\

Mnih, V., Kavukcuoglu, K., Silver, D., Rusu, A. A., Veness, J., Bellemare, M. G., Graves, A., Riedmiller, M., Fidjeland, A. K., Ostrovski, G., Petersen, S., Beattie, C., Sadik, A., Antonoglou, I., King, H., Kumaran, D., Wierstra, D., Legg, S., \& Hassabis, D. (2015). Human-level control through deep reinforcement learning. Nature, 518, 529-533.\\

Oborn, E., Barrett, M., \& Dawson, S. (2013). Distributed leadership in policy formulation: A sociomaterial perspective. Organization Studies, 34(2), 253-276.\\

Orlikowski, W. J. (2000). Using technology and constituting structures: A practice lens for studying technology in organizations. Organization Science, 11(4), 404-428.\\

Orlikowski, W. J. (2007). Sociomaterial practices: Exploring technology at work. Organization Studies, 28(9), 1435-1448.\\

Orlikowski, W. J., \& Scott, S. V. (2008). Sociomateriality: Challenging the separation of technology, work and organization. Academy of Management Annals, 2(1), 433-474.\\

Pentland, B. T., Haerem, T., \& Hillison, D. (2010). Comparing organizational routines as recurrent patterns of action. Organization Studies, 31(7), 917-940.\\

Powell, W. W. (1990). Neither market nor hierarchy: Network forms of organization. Research in Organizational Behavior, 12, 295-336.\\

Raisch, S., \& Krakowski, S. (2021). Artificial intelligence and management: The automation-augmentation paradox. Academy of Management Review, 46(1), 192-210.\\

Ratner, H., \& Plotnikof, M. (2022). Technology and dis/organization: Digital data infrastructures as partial connections. Organization Studies, 43(7), 1049-1067.\\

Riedl, C., De Cremer, D., Lucarelli, G., \& Antoine-Souklaye, E. (2025). The potential and challenges of AI for collective intelligence. Collective Intelligence, 4(1).\\

Shrestha, Y. R., Ben-Menahem, S. M., \& von Krogh, G. (2019). Organizational decision-making structures in the age of artificial intelligence. California Management Review, 61(4), 66-83.\\

Simon, H. A. (1947). Administrative Behavior. Macmillan.\\

Star, S. L., \& Griesemer, J. R. (1989). Institutional ecology, translations, and boundary objects. Social Studies of Science, 19(3), 387-420.\\

Stark, D., \& Vanden Broeck, P. (2024). Principles of algorithmic management. Organization Theory, 5(2).\\

Steinberg, M. (2022). From automobile capitalism to platform capitalism: Toyotism as a prehistory of digital platforms. Organization Studies, 43(7), 1069-1090.\\

Symon, G., \& Pritchard, K. (2015). Performing the responsive and committed employee through the sociomaterial mangle of connection. Organization Studies, 36(2), 241-263.\\

Thompson, J. D. (1967). Organizations in Action. McGraw-Hill.\\

Tsoukas, H. (1996). The firm as a distributed knowledge system: A constructionist approach. Strategic Management Journal, 17(S2), 11-25.\\

Tsvetkova, M., Yasseri, T., Pescetelli, N., \& Werner, T. (2024). A new sociology of humans and machines. Nature Human Behaviour, 8, 1864-1876.\\

Weick, K. E. (1979). The Social Psychology of Organizing. Addison-Wesley.\\

Williamson, O. E. (1985). The Economic Institutions of Capitalism. Free Press.\\

Woolley, A. W., Chabris, C. F., Pentland, A., Hashmi, N., \& Malone, T. W. (2010). Evidence for a collective intelligence factor in the performance of human groups. Science, 330(6004), 686-688.\\

\end{hangparas}

\appendix

\section{Appendix A. Additional Simulation Evidence}

\begin{table}[htbp]
\centering
\small
\caption{Descriptive simulation results by organization form}
\begin{tabularx}{\textwidth}{@{}YZZZZZZ@{}}
\toprule
Organization form & Family & Quality & Success & Cost & CTC share & Collective efficiency \\
\midrule
Adaptive meta-organization & Agent-native & 57.83 & 63.85\% & 5.30 & 19.12\% & 24.23 \\
Blackboard memory & Agent-native & 55.58 & 59.35\% & 7.48 & 33.71\% & 20.41 \\
Market contract & Hybrid & 47.17 & 45.78\% & 10.30 & 60.97\% & 12.83 \\
Single expert & Compact & 43.84 & 39.40\% & 4.04 & 0.00\% & 12.80 \\
Pipeline SOP & Human-imitation & 37.67 & 33.79\% & 9.04 & 64.31\% & 8.52 \\
Hierarchy manager & Human-imitation & 32.35 & 28.34\% & 21.38 & 78.79\% & 4.89 \\
Committee debate & Human-imitation & 1.74 & 15.24\% & 70.84 & 92.92\% & 0.11 \\
\bottomrule
\end{tabularx}
\end{table}

\begin{table}[htbp]
\centering
\small
\caption{Task winner distribution}
\begin{tabularx}{\textwidth}{@{}YZZZ@{}}
\toprule
Winning organization form & Family & Tasks won & Share of tasks won \\
\midrule
Adaptive meta-organization & Agent-native & 5,536 & 69.20\% \\
Blackboard memory & Agent-native & 2,020 & 25.25\% \\
Single expert & Compact & 365 & 4.56\% \\
Market contract & Hybrid & 76 & 0.95\% \\
Pipeline SOP & Human-imitation & 3 & 0.04\% \\
\bottomrule
\end{tabularx}
\end{table}

\section{Appendix B. Prompt Robustness and Prompt-Free Organization Selection}

Prompt robustness analysis addresses the possibility that the observed patterns are artifacts of one prompt. OpenAI runs are stable across tested styles. Qwen2 open-source runs are more sensitive, which supports the claim that prompt and protocol dependence is itself an organizational property of agentic AI.

\begin{table}[htbp]
\centering
\small
\caption{Qwen2 prompt robustness}
\begin{tabularx}{\textwidth}{@{}YZZZZZZ@{}}
\toprule
Model & Organization & Prompt styles & Quality CV & Efficiency CV & Efficiency range & Mean CTC \\
\midrule
Qwen2 7B instruct & Single & 4 & 24.32\% & 37.87\% & 17.22 to 63.34 & 1.06 \\
Qwen2 7B instruct & Adaptive & 4 & 24.03\% & 35.53\% & 12.18 to 38.95 & 7.42 \\
\bottomrule
\end{tabularx}
\end{table}

\begin{table}[htbp]
\centering
\small
\caption{Prompt-free fitted DQN organization selector}
\begin{tabularx}{\textwidth}{@{}YZZZZ@{}}
\toprule
Policy & Mean collective efficiency & Mean quality & Success rate & Dominant selected form \\
\midrule
Hand-coded adaptive meta-organization & 24.21 & 57.87 & 63.95\% & Adaptive \\
Prompt-free DQN selector & 24.14 & 57.77 & 63.65\% & Adaptive \\
DQN without adaptive action & 20.33 & 55.21 & 58.05\% & Blackboard \\
Best fixed no-adaptive baseline & 20.31 & 55.49 & 58.15\% & Blackboard \\
Single expert baseline & 12.70 & 43.71 & 39.95\% & Single \\
\bottomrule
\end{tabularx}
\end{table}

The DQN selector receives task features and chooses an organization action to maximize collective efficiency. It does not receive natural-language prompts. The fitted Q-learning loss is

\[
L(\phi)=
\left[r(T,O)-Q_\phi(x_T,O)\right]^2,
\]

where \(x_T\) is the task feature vector, \(O\) is the organization action, and \(r(T,O)\) is collective efficiency. The result separates the organizational mechanism from prompt wording. A prompt-free selector learns a policy close to the hand-coded adaptive form when that action is available and selects blackboard memory when it is not.

\end{document}